\documentclass[useAMS,usenatbib]{mn2e}

\usepackage{graphicx}
\usepackage[dvips]{color}

\def\be{\begin{equation}}
\def\ee{\end{equation}}
\def\ba{\begin{eqnarray}}
\def\ea{\end{eqnarray}}

\def\12{{1\over 2}}

\def\msun{M_\odot}

\def\etal{{\it et~al.~}}
\def\ltsima{$\; \buildrel < \over \sim \;$}
\def\simlt{\lower.5ex\hbox{\ltsima}}
\def\gtsima{$\; \buildrel > \over \sim \;$}
\def\simgt{\lower.5ex\hbox{\gtsima}}

\title[Particle decay and 21 cm from first minihaloes]{\bf Particle 
decay and 21 cm absorption from first minihaloes}
\author[E. O. Vasiliev and Yu. A. Shchekinov]
       {E. O. Vasiliev$^{1,2}$\thanks{E-mail:eugstar@mail.ru}
and  Yu. A. Shchekinov$^{2,3}$\thanks{E-mail:yus@sfedu.ru}\\
$^1$Institute of Physics, Southern Federal University, 
Stachki Ave. 194, Rostov-on-Don, 344090 Russia \\
$^2$Department of Physics, Southern Federal University, 
Sorge St. 5, Rostov-on-Don, 344090 Russia\\ 
$^3$Special Astrophysical Observatory RAS, Nizhny Arkhyz, 369167 Russia
}
\begin{document}
\date{Accepted 2006 December 15.
      Received 2006 April 1;
      in original form 2006 April 1}
\pagerange{\pageref{firstpage}--\pageref{lastpage}}
\pubyear{2006}
\maketitle

\label{firstpage}

\begin{abstract}
We consider the influence of decaying dark matter (DM) particles on the characteristics of 21 cm absorption 
in spectra of distant radio-loud sources -- ``21 cm forest'' -- from minihaloes with masses  
$M=10^5-10^7\msun$ virialized at $z_{vir} = 10$. We use 1D self-consistent hydrodynamic description
to study evolution of minihaloes, and follow up their absorption characteristics from turnaround to 
virialization. We find that in the presence of decaying dark matter both thermal and dynamical evolution 
of minihaloes demonstrate significant deviation from those in the model without dark matter
decay (standard recombination). We show that optical depth in 21 cm line is strongly 
suppressed in the presence of decaying particles: for $M=10^5-10^6\msun$ decaying dark matter with the energy rate 
deposited in baryonic gas $\xi_{L} = 0.59\times 10^{-25}$~s$^{-1}$ -- the current upper limit of the energy 
deposit -- decreases the optical depth and the equivalent width by an order of magnitude compared to the standard 
recombination. Thus additional ionization and heating from decaying DM particles almost ``erases'' absorption 
features from minihaloes with $M=10^5-10^6\msun$ for $\xi \simgt 0.3\xi_L$, which consequently considerably
decreases the number of strong absorptions: for example, the number of absorptions with the equivalent width 
$W_\nu^{obs} \simgt 0.3$~kHz at $z\simeq10$ decreases more than 2.5 times for $\xi/\xi_{L} = 0.3$ and 
$\simgt$4.5 times for $\xi/\xi_{L} = 1$. We argue that ``21 cm forest'' absorptions might be a powerful probe 
of the presence of decaying dark matter in the early Universe.
\end{abstract}


\section{Introduction}

\noindent

A way to study pre-reionization history of the intergalactic medium lies in observation of the redshifted 
21 cm line produced in the hyperfine transition of neutral hydrogen \citep{hogan,scott,mmr}. There are at least 
three techniques to use the 21 cm line for studying this epoch: observations of emission or absorption of the 
neutral intergalactic medium (IGM) against the cosmic microwave background (CMB) -- the 21 cm global signal
\citep{shaver99,sethiHI}, statistical studies of the angular distribution of the intensity in the 21 cm line 
-- the 21 cm fluctuations \citep{tozzi,iliev02}, and measurements of absorption along the line of sight to a distant
radio-loud quasar -- the "21 cm forest"  \citep{carilli02,furla02}. The first two methods provide mapping of the
3D space-redshift features of the IGM, whereas the latter works only along line of sight, i.e. provides a 1D probe. 
A physical distance corresponded to the angular resolution of existing and future radiotelescopes, such as e.g. 
LOFAR\footnote{http://www.lofar.org/index.htm}, is about one megaparsec. Therefore only huge structures, such as
for example, HII regions formed by stellar clusters and quasars, the large scale cosmic web structure, are 
possible to detect. Instead, by measuring the absorption features along redshift we can resolve small scale structure, 
e.g. individual minihaloes, galaxies, stellar HII regions \citep{kumar,nathHI,furla02,furla06,ciardi21,mack11}. 

Ionizing photons produced by first stellar objects change gas thermal state in minihaloes and their
vicinity, and as a consequence, affect the 21 cm absorption features \citep{ciardi21,ferrara11}. 
Similar effect can be expected from other possible sources of ionizing photons, such as decaying dark matter (DM) 
particles \citep{sciama82,scott,sethiddm,kamiondecays,hh,kasuya04a,kasuya04b,pierpaoli,belikov}, annihilation 
of dark matter \citep{chuzhoy07,nusser07,leptonddm,silk21ddm} and ultra-high energy cosmic rays produced in 
decay of superheavy dark matter particles with masses $M_X \simgt 10^{12}$~GeV \citep{berez,birkel,kuzmin,DN,dnnn}. 
Previous studies have shown that ionizing background produced by these sources can affect thermal and ionization
evolution of the intergalactic medium \citep{dodelson,biermannH2ddm,sh04,vas06,mapelli} and influence through it
the 21 cm global signal and fluctuations \citep{furl06,sh07,chuzhoy07,nusser07,leptonddm,silk21ddm,natarajan}. 
Recent studies strongly constrain properties of such ionizing radiation sources
\citep{kamion07,anihreion,delopeddm,kamion10,zhangddm,galli11anih,wmapanih}, though do not fully exclude them. 
Therefore, in this paper we focus on how the 21 cm absorption -- ``21 cm forest'' -- depends on the decaying dark 
matter parameters. 

The 21 cm forest is supposed to be a signal from numerous minihaloes along line of sight being at different 
evolutionary states from the turnaround to the virialization, with the absorption features depending on the 
evolutionary state \citep{meiksin11,vs12}. In several previous studies minihaloes were considered as static 
objects with fixed dark matter and baryonic profiles \citep{furla02,furla06}. More recently accretion has been
added \citep{ferrara11}, and further, the evolutionary effects during the formation of minihaloes have been 
taken into account \citep{meiksin11,vs12}. 

Another aspect important when characteristics of the 21 cm forest are concerned relates to spatial 
distribution of matter in minihaloes. The dark matter profile is mostly assumed cuspy \citep{nfw}. However, 
according to the observations of local dwarf galaxies the dark matter profile is flatter \citep{burkert}. 
Recently in several theoretical studies it has been pointed out that the first protogalaxies may 
also have a flat dark matter profile \citep{mashch,tonini,dor06}. It seems obvious that changes in dark matter 
profiles result in corresponding changes of the radial distribution of baryonic component inside minihaloes, and
consequently, in the 21 cm absorption properties.

In this paper we study the effects of decaying dark matter particles on absorption features of non-static (evolving)
minihaloes with a flat dark matter profile. We assume a $\Lambda$CDM cosmology with the parameters $(\Omega_0,\Omega_{\Lambda},\Omega_m,\Omega_b,h)=(1.0,0.76,0.24,0.041,0.73)$ \citep{wmap}.

\section{Evolution of minihalo}

To model the evolution of minihaloes we use a one-dimensional Lagrangian code. 
For dark matter we follow the description given by \citet{ripa}. Namely, the 
dark matter, $M_{DM} = \Omega_{DM} M_{halo}/\Omega_M$, is assumed to be enclosed 
within a certain truncation radius $R_{tr}$, inside which the dark matter profile 
is a truncated isothermal sphere with a flat core of radius $R_{core}$. The parameter 
$\eta = R_{core} /R_{vir}$ is taken 0.1 for all simulations. Such a description is 
used to mimic the evolution of a simple top-hat fluctuation \citep[e.g.][]{padma}.
In order to describe time evolution of the truncation radius we make use of the 
approximation formula for a top-hat density perturbation in the form proposed by \citet{t97}.

For simulations of baryonic component we use a 1D Lagrangian scheme similar to that 
described by \citet{1d}. As a standard resolution we used 1000 zones over the 
computational domain, and found a reasonable convergence.

Chemical and ionization composition include a standard set of species: 
H, H$^+$, H$^-$, He, He$^+$, He$^{++}$, H$_2$, H$_2^+$, D, D$^+$, D$^-$, HD, HD$^+$ and $e$.
The corresponding reaction rates are taken from \citep{galli98,stancil98}. 
Energy equation includes radiative losses typical in primordial plasma:
Compton cooling, recombination and bremsstrahlung radiation, collisional 
excitation of HI \citep{cen92}, H$_2$ \citep{galli98} and HD \citep{flower00,lipovka05}. 

We start our one-dimensional simulations at redshift $z = 100$. The initial parameters: 
gas temperature, chemical composition of gas and other quantities -- we have taken from 
simple one-zone calculations began at $z = 1000$ with typical values at the end of 
recombination: $T_{gas} = T_{CMB}$, $x[{\rm H}] = 0.9328, x[{\rm H^+}] = 0.0672, 
x[{\rm D}] = 2.3\times 10^{-5}, x[{\rm D^+}] = 1.68\times 10^{-6}$ \citep[see references 
and details in][Table 2]{ripa}.

\section{Ionization and thermal history}

The contribution from decaying dark matter is described by the corresponding ionization and heating 
terms into equations for the ionization and thermal evolution. The ionization rate due 
to presence of additional ionization sources from decaying particles is written as 
\citep{kamiondecays}
\be
I_e(z) = \chi_i f_x \Gamma_X { m_p c^2 \over h \nu_c}
\ee
where $\chi_i$ is the energy fraction deposited into ionization \citep{shull}, $m_p$ is the 
proton mass, $f_x =\Omega_X(z)/\Omega_b(z)$, $\Omega_b(z)$, the baryon density parameter, 
$\Omega_X(z)$, the fractional abundance of decaying particles, $\Gamma_X$ is the decay rate, 
$h\nu_c$, the energy of Ly-c photons. In general, the ionization history depends on the 
energy rate deposited in baryonic gas, $\xi = \chi_i f_x \Gamma_X$ \citep{kamiondecays}.

The heating rate produced by ionizing photons from decaying dark matter particles can be written 
in the form \citep{kamiondecays}
\be
K = \chi_h m_p c^2 f_x \Gamma_X
\ee
where $\chi_h$ is the energy fraction depositing into heating \citep{shull}. By order of 
magnitude $\chi_i\sim\chi_h\sim 1/3$ for the conditions we are interested in. 

Using the CMB datasets \citet{kamion07} have constrained the energy rate of the radiatively decaying 
dark matter as the upper limit $\xi_L  \simlt 1.7\times 10^{-25}$~s$^{-1}$. Extending analysis with data 
of Type Ia supernova, Ly$\alpha$ forest, large scale structure and weak lensing observations have lead 
\citet{delopeddm} to stronger constraints: $\xi_L \simlt 0.59\times 10^{-25}$~s$^{-1}$.
It is worth noting that all datasets favour long-living decaying dark matter particles with the lifetime 
$\Gamma_X^{-1} \simgt 100$~Gyr \citep{delopeddm}. Further improvement is expected from the Planck satellite.
In this paper we consider the decaying dark matter with $\xi \le \xi_{L} = 0.59\times 10^{-25}$~s$^{-1}$ and 
compare results with the standard recombination scenario, i.e. $\xi=0$.

\section{Spin temperature}

The two processes: atomic collisions and scattering of ultraviolet (UV) photons, couple
the HI spin temperature and the gas kinetic temperature \citep{field,wout}
\be
T_s = {T_{CMB} + y_a T_k + y_c T_k \over 1 + y_a + y_c}
\label{ts}
\ee
here $T_{CMB}$ is the CMB temperature, $y_c, y_a$ are the functions 
determined by the collisional excitations and the intensity of the UV 
resonant photons
\be
 y_a = {P_{10}T_\ast \over A_{10} T_k}, \hspace{1cm}
 y_c = {C_{10}T_\ast \over A_{10} T_k}
\ee
$T_\ast = 0.0682$~K is the hyperfine energy splitting,
$A_{10} = 2.87\times 10^{-15}$~s$^{-1}$ is the spontaneous emission
rate of the hyperfine transition,  $C_{10} = k_{10}n_H + \gamma_e n_e$
is the collisional de-excitation rate by hydrogen atoms and electrons,
the photon induced de-excitation rate is negligible, for $k_{10}$ we use the approximation
by \citet{kuhlen}, for $\gamma_e$ we take the approximation from \citet{liszt},
$P_{10}$ is the indirect de-excitation rate, which is related to the total Ly$\alpha$ scattering
rate $P_a$ \citep{field}
\be
 P_{10} = 4 P_a / 27,
\ee
where $P_a = \int c n_\nu \sigma_\nu d\nu$, 
$n_\nu$ is the number density of photons 
per unit frequency range, $\sigma(\nu)$ is the
cross-section for Ly$\alpha$ scattering \citep{mmr}.

Additional ionizations produced by decaying dark matter deposit into Ly$\alpha$ photon 
budget due to recombination. In our calculations we do not take into account the 
UV pumping, firstly because such deposit is minor in comparison with the atomic 
collisions for the decaying dark matter parameters and redshifts considered here 
\citep[see Figure 2 in][]{sh07}. Secondly, contribution from Ly$\alpha$ background 
produced by star-forming minihaloes is small due to the rarity of first stellar 
objects at redshifts $z=10-15$. Moreover, it can be readily shown that in order to 
compete contribution from atomic and electron collisions to HI spin temperature UV 
energy flux has to be unacceptably high: $F_{\rm Ly\alpha}\simgt 0.1$~erg~s$^{-1}$~cm$^{-2}$ 
for gas with $T\sim 200$~K and $n\sim 1$~cm$^{-3}$ at $z\sim 10$.

\section{Results}

\noindent

Stellar and quasi-stellar sources of ionizing radiation begin to form at redshifts $z<20$.  
They heat the surrounding gas through photoionization, which then emits in 21 cm with a 
patchy, spot-like distribution on the sky. Instead, the decaying particles illuminate 
and heat the IGM homogeneously by X-ray photons normally produced in decays 
\citep{kamiondecays}. Such photons have mean free path length much longer than the typical 
distance between minihaloes, $d\sim 30(1+z/20)$~kpc. Therefore in minihaloes the ionization 
and heating rates due to decays of dark matter particles can be taken equal to the rates in the IGM 
background.

\subsection{Dynamics of minihaloes}

Heating from decaying dark matter weakens the accretion rate of baryons on to the minihaloes. Therefore, 
in comparison with the evolution of minihaloes in the standard scenario one can expect smaller baryon mass and higher 
temperature within the virial radius of a minihalo. The influence of decaying particles depends  
on the minihalo mass and the rate of the decay energy deposited in gas, $\xi$. 

Here we consider evolution of minihaloes within the mass range $M=10^5-10^7\msun$ virialized 
at $z_v=10$. Our choice of the redshift of virialization is caused by the fact that at lower 
redshifts the effects of reionization begin to play dominant role and thus contaminated contribution 
from decaying dark matter. On the other hand, at higher redshifts the number of bright radio objects which can 
serve as background sources where minihaloes can imprint becomes low. In the mass range we focus on the
three characteristic masses: $M=10^5$, $10^6$ and $10^7~M_\odot$ for the following reasons. 
It has been found that inside minihaloes with $M \simeq 2\times 10^6\msun$ the star formation can potentially 
occur at $z\simeq 10$ \citep[see also][]{ripa}, in the sense that such minihaloes collapse at $z_{vir}=10$, i.e. the 
gas density in the most inner shell reaches around $10^8$~cm$^{-3}$ and grows up further. Minihaloes with lower masses 
cannot cool efficiently and form stars at $z \sim 10$. Therefore the first mass considered here $M=10^5\msun$
represents in general the low-mass end of non-starforming minihaloes. The second value $M=10^6\msun$ cannot also
collapse at $z \sim 10$, but it is close to the critical mass $M \simeq 2\times 10^6\msun$, and represents
qualitatively correct characteristics of 21 cm absorptions in the range of masses of marginally star-forming
minihaloes. The third value is $M=10^7\msun$: such minihaloes collapse (and therefore can form stars) even before 
the virialization redshift, namely, at $z\simeq 11$. Thus, we study evolution and absorption characteristics of the
minihaloes, which from one side can remain dark after virialization ($M=10^5-10^6\msun$) and from the other 
form first stars ($M=10^7\msun$). 

\begin{figure*}
\includegraphics[width=75mm]{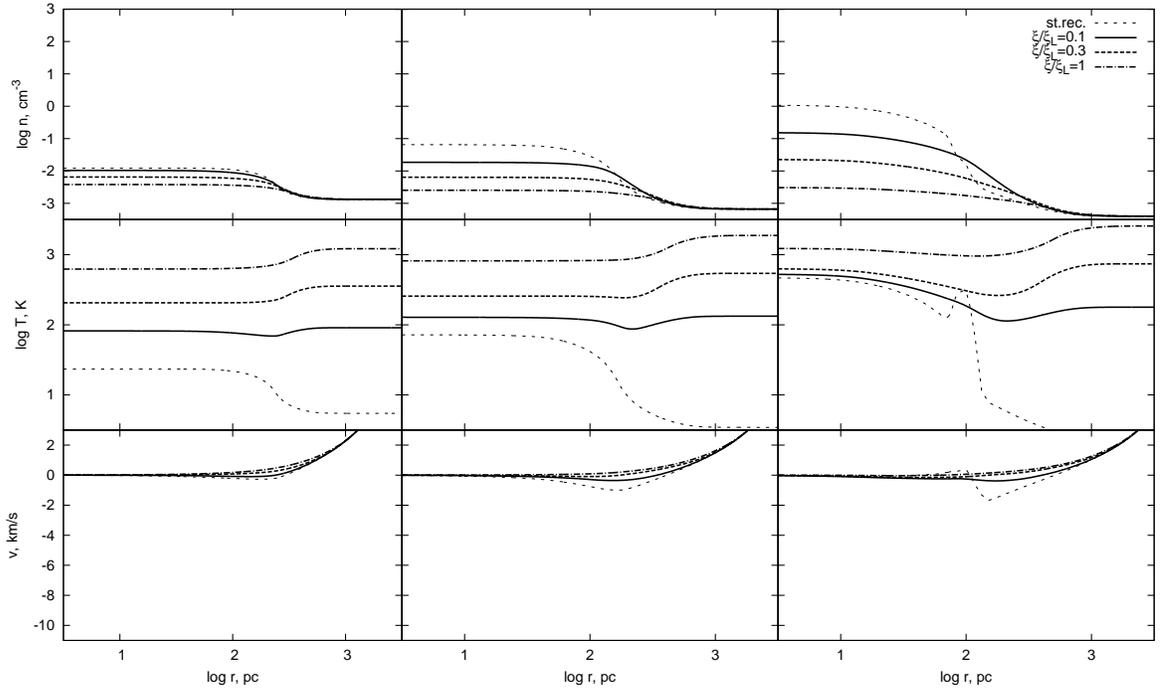}
\caption{
The radial density (upper), temperature (middle) and velocity (lower panels) profiles
of a halo $M=10^5\msun$ virialized at $z_{vir} = 10$ are shown at redshifts $z = 15.5, 12$ 
and 10 (from left to right panels, correspondingly) in the standard recombination model (dots)
and in the presence of decaying dark matter with $\xi/\xi_{L} = 0.1, 0.3, 1$ (solid, dash and
dot-dashed lines, correspondingly), where $\xi_{L} = 0.59\times 10^{-25}$~s$^{-1}$. Note 
that  in the presence of decaying dark matter gas almost ceases collapse into the minihalo 
potential well.
}
\label{fig1}
\end{figure*}

\begin{figure*}
\includegraphics[width=75mm]{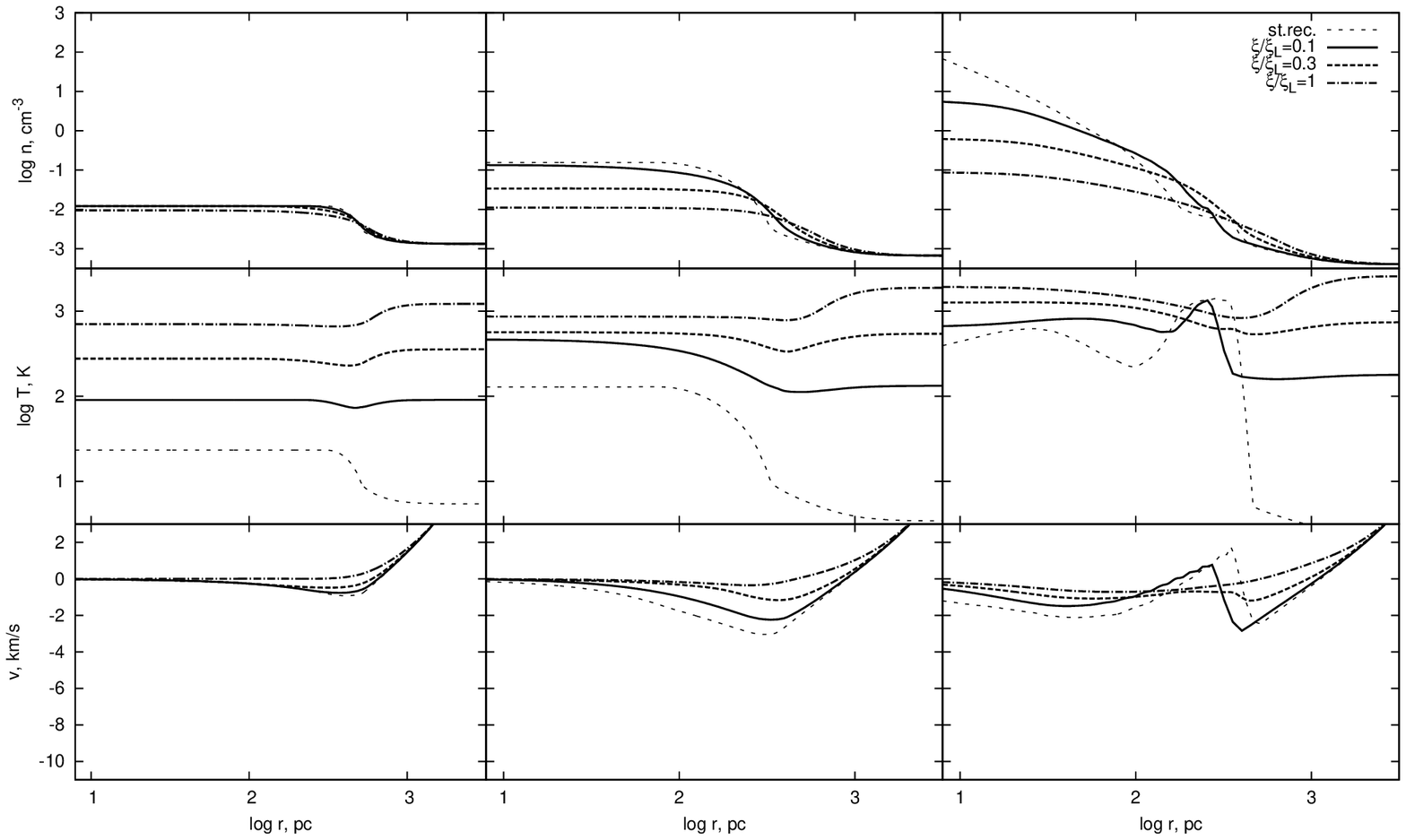}
\caption{
The radial density (upper), temperature (middle) and velocity (lower panels) profiles
of a halo $M=10^6\msun$ virialized at $z_{vir} = 10$ are shown at redshifts $z = 15.5, 12$ 
and 10 (from left to right panels, correspondingly) in the standard recombination model (dots)
and in the presence of decaying dark matter with $\xi/\xi_{L} = 0.1, 0.3, 1$ (solid, dash and
dot-dashed lines, correspondingly), where $\xi_{L} = 0.59\times 10^{-25}$~s$^{-1}$.
Note the absence of the accretion (virial) shock for $\xi/\xi_{L} \simgt 0.3$ (right lower panel), 
whereas for the lower $\xi/\xi_{L}$ the profiles tend to those in the standard recombination model.
}
\label{fig2}
\end{figure*}

\begin{figure*}
\includegraphics[width=75mm]{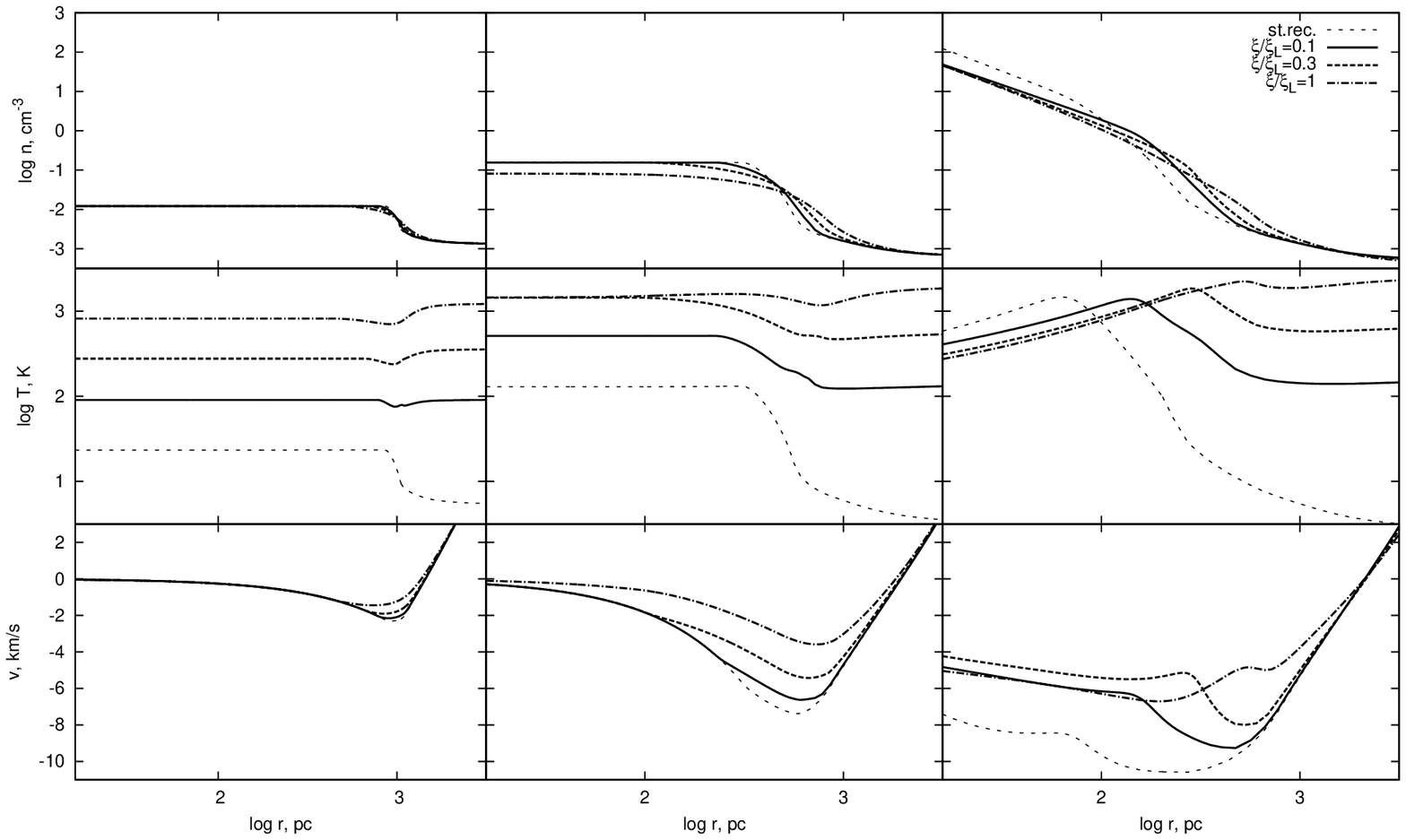}
\caption{
The radial density (upper), temperature (middle) and velocity (lower panels) profiles
of a halo $M=10^7\msun$ virialized at $z_{vir} = 10$ are shown at redshifts $z = 15.5, 12$ 
and 11 (from left to right panels, correspondingly) in the standard recombination model (dots)
and in the presence of decaying dark matter with $\xi/\xi_{L} = 0.1, 0.3, 1$ (solid, dash and
dot-dashed lines, correspondingly), where $\xi_{L} = 0.59\times 10^{-25}$~s$^{-1}$. Note that
the velocity profiles demonstrate a stable gas collapse in the potential well of the dark halo. 
The density and velocity profiles for all considered $\xi/\xi_{L}$ are close to those in the 
standard scenario, though the temperature profiles $\xi/\xi_{L}$ demonstrate appreciable difference.
}
\label{fig3}
\end{figure*}

\subsubsection{Evolution of minihaloes with $M=10^5\msun$ and $10^6\msun$.}

Figures~\ref{fig1}-\ref{fig2} show radial profiles of number density (upper), temperature (middle) 
and velocity (lower panels) in minihaloes $M=10^5, 10^6\msun$ virialized at $z_{vir} = 10$, 
for the three redshifts $z = 15.5, 12$ and 10 (from left to right panels, correspondingly).  

For $M=10^5\msun$ the virial temperature about 500~K is close to the one 
in the IGM heated by decaying DM particles with $\xi/\xi_{L} \simgt 0.1$. At the turnaround state
(left panels) temperature profiles differ significantly from the standard one: namely, they 
show the internal and the external temperatures practically equal due to additional heating from 
decaying DM particles. Such a high background temperature in the presence of decaying 
particles can cause stopping gas collapse in the minihalo potential well during further evolution. 

At $z=12$ (middle panels of Figure~\ref{fig1}) the standard model shows obvious gas accretion 
resulting in an increase of density and temperature. At the same time in models with
$\xi \simgt 0.3\xi_{L}$ gas velocities are non-negative at any radius, so that the 
baryon content of minihalo does not grow. The density within the plateau $r\simlt 100$~pc 
for $\xi = \xi_{L}$ is an order of magnitude lower than in the standard recombination -- 
this is clearly seen on the lower panel of Figure~\ref{fig1}. At the virialization this difference 
in density increases: for low ionization rates $\xi \simlt 0.3\xi_{L}$ the density grows
several times compared to that at $z=12$, while for higher $\xi = \xi_{L}$ it stays nearly the 
same. As a result for high ionization rates the deviation from the standard model reaches
two orders of magnitude. Practically in the whole range of ionization rates temperature profiles 
within $r\simlt 100$~pc do not deviate significantly from each other, showing a dominance of the 
decaying DM in heating. The most appreciable difference between the standard recombination 
and the models with decaying dark matter is the absence of a significant accretion shock wave around 
$r\sim 100$~pc. In the standard case such a wave is obviously seen on the right lower panel 
of Figure~\ref{fig1}. Thus the presence of decaying dark matter prevents collecting baryonic mass 
in the minihaloes with $M\simeq 10^5\msun$.

For $\xi \simgt 0.3\xi_{L}$ the evolution of minihaloes with $M=10^6\msun$ is similar to that with 
$M=10^5\msun$  (the dash and dash-dotted lines in Figure~\ref{fig2}). Whereas for lower 
rate $\xi$ evolution of $M=10^6\msun$ looks closer to the standard recombination model. 
Note, in particular, the presence of the virial shock for $\xi = 0.1\xi_{L}$ (solid lines at the lower 
row of panels in Figure~\ref{fig2}) similar to that in the standard case (dotted line). 

\subsubsection{Evolution of minihaloes with $M=10^7\msun$.}

\begin{figure*}
\includegraphics[width=75mm]{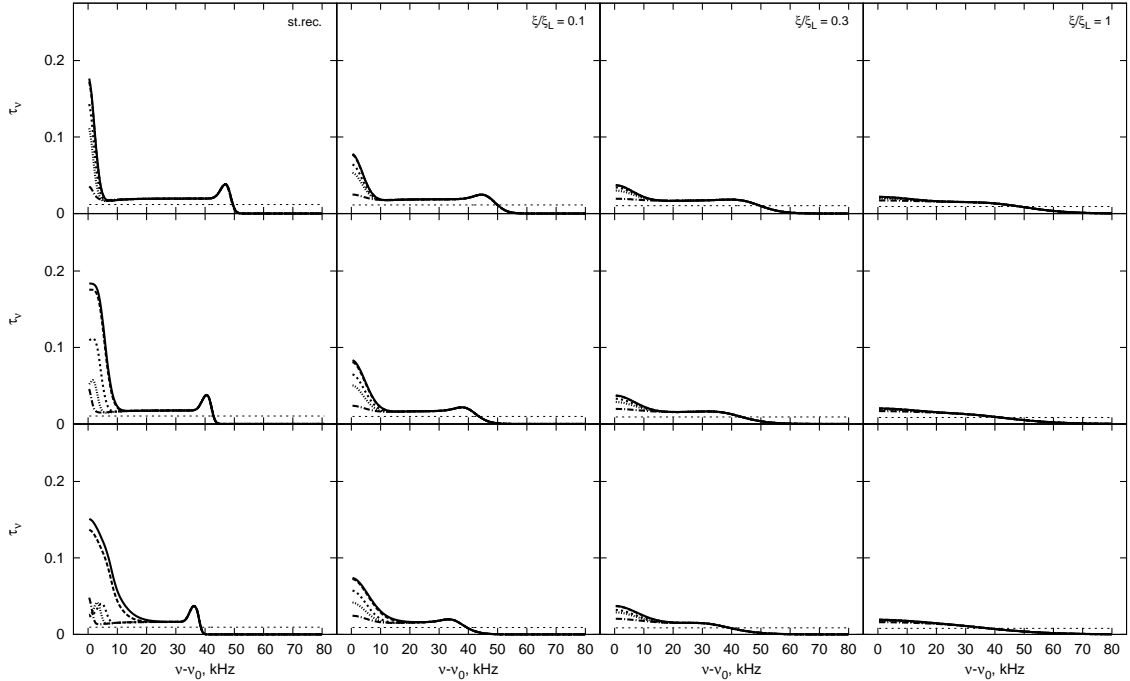}
\caption{
Optical depth at the impact parameters $\alpha=0.1,\  0.3,\  1,\  1.5,\ 3r/r_{vir}$ (solid, dash,
dot, short dash and dot-dash lines, correspondingly) for a halo $M=10^5\msun$  
virialized at $z_{vir} = 10$, at redshifts $z = 15.5, 12$ and 10 (from top to bottom panels, 
correspondingly) in the standard recombination model and in the presence of decaying dark matter with 
$\xi/\xi_{L} = 0.1, 0.3, 1$ (from left to right, correspondingly), where 
$\xi_{L} = 0.59\times 10^{-25}$~s$^{-1}$.
}
\label{fig4}
\end{figure*}

For $M=10^7\msun$ the density profiles\footnote{As for minihaloes with $M=10^7\msun$ the collapse (i.e. the time 
when gas density in the most inner shell reaches $10^8$~cm$^{-3}$ and grows further) occurs earlier than formal
virialization, we present the results for not later than the collapse redshift, i.e. for $z=11$.} for all $\xi$ 
considered here are close to those in the standard recombination scenario (Figure~\ref{fig3}). Instead, temperature 
and velocity profiles demonstrate considerable differences. These differences can be explained by the accretion of
warmer intergalactic gas. At the turnaround temperature inside the halo is almost equal to the background value 
(middle left panel), which is higher than the standard background temperature. At this stage gas has already begun
collapsing into the potential well of the dark halo, providing the velocity profiles almost coincided for all
$\xi\simlt\xi_L$ (lower left panel). In further accretion of warm intergalactic gas the differences between models
start revealing at $z=12$. Temperature inside the minihalo ($r\simlt 500$~pc) becomes considerably higher for the 
models with decaying particles. The absolute value of the velocity decreases with an increase of the rate $\xi$, 
i.e. baryons collapse slower. At the collapse redshift, $z\simeq 11$, differences in the temperature and the velocity
profiles for different rates $\xi$ remain significant in the whole radial range. In all models there the virial shock
wave does not appear, the velocity profiles demonstrate stable collapse of gas in the potential well of the dark halo.
One can note that the profiles for the models with decaying particles are close each other and to the profile for the
standard reionization within $r\simlt 200$~pc which corresponds to $r \simlt 0.5 r_{vir}$. 

The evolutionary differences for minihaloes in the mass range $M=10^5 - 10^7\msun$ in the decaying dark matter 
models are expected to become seen in the HI 21 cm absorption characteristics. In the following sections we 
consider how the differences manifest in optical depth and equivalent width of the 21 cm absorption.

\subsection{Optical depth}

The optical depth at a frequency $\nu$ along a line of sight is

\ba
\label{optd} 
 \tau_\nu = {3h_p c^3 A_{10}\over 32 \pi k \nu_0^2} \int_{-\infty}^{\infty}
   dx {n_{HI}(r) \over \sqrt{\pi} b^2(r) T_s(r)} \\ 
   \nonumber
   {\rm exp}\left[{-{[v(\nu) - v_l(r)]^2\over b^2(r)}}\right]
\ea
where $r^2 = (\alpha r_{vir})^2 + x^2$, $\alpha=r/r_{vir}$ is the impact parameter (a line of sight crosses the 
minihalo at a distance $\alpha r_{vir}$ from the centre of the minihalo), $v(\nu)=c(\nu-\nu_0)/\nu_0$, $v_l(r)$ 
is the infall velocity projected on the line of sight, $b^2 = 2kT_k(r)/m_p$ is the Doppler parameter of gas.

\begin{figure*}
\includegraphics[width=75mm]{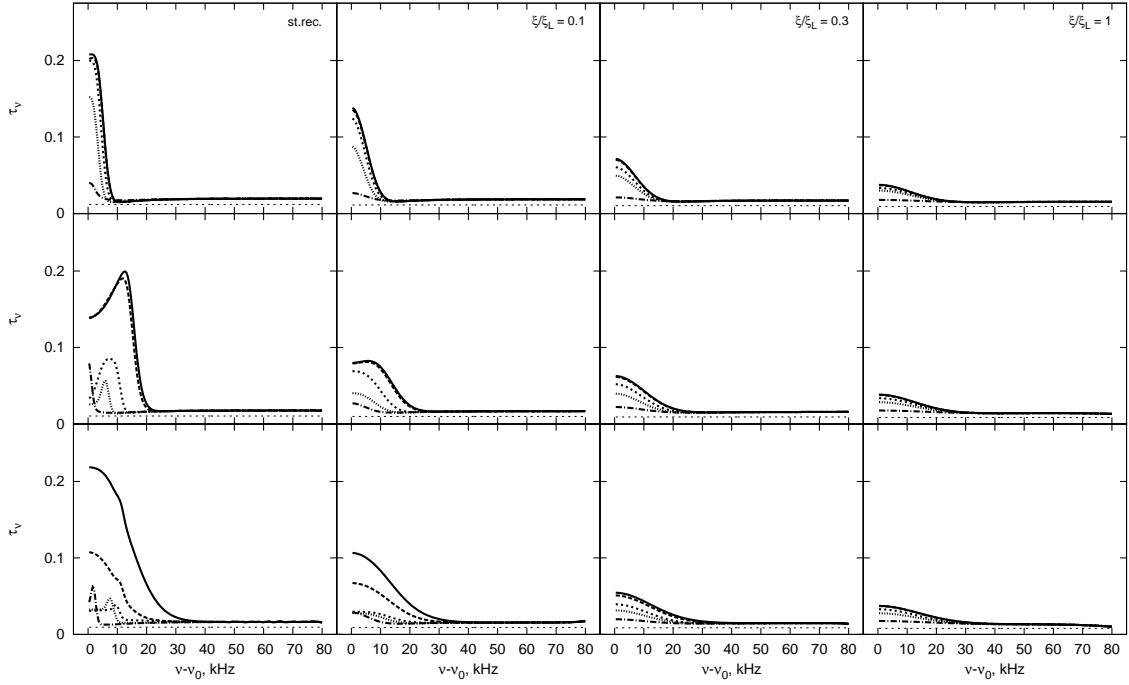}
\caption{
The same as in Figure~\ref{fig4}, but for a halo with $M=10^6\msun$: $\alpha=0.1,\  0.3,\  1,\  1.5,\ 3r/r_{vir}$ (solid, dash, dot, short dash and dot-dash lines, correspondingly), at redshifts $z = 15.5, 12$ and 10 (from top 
to bottom). 
}
\label{fig5}
\end{figure*}

\begin{figure*}
\includegraphics[width=75mm]{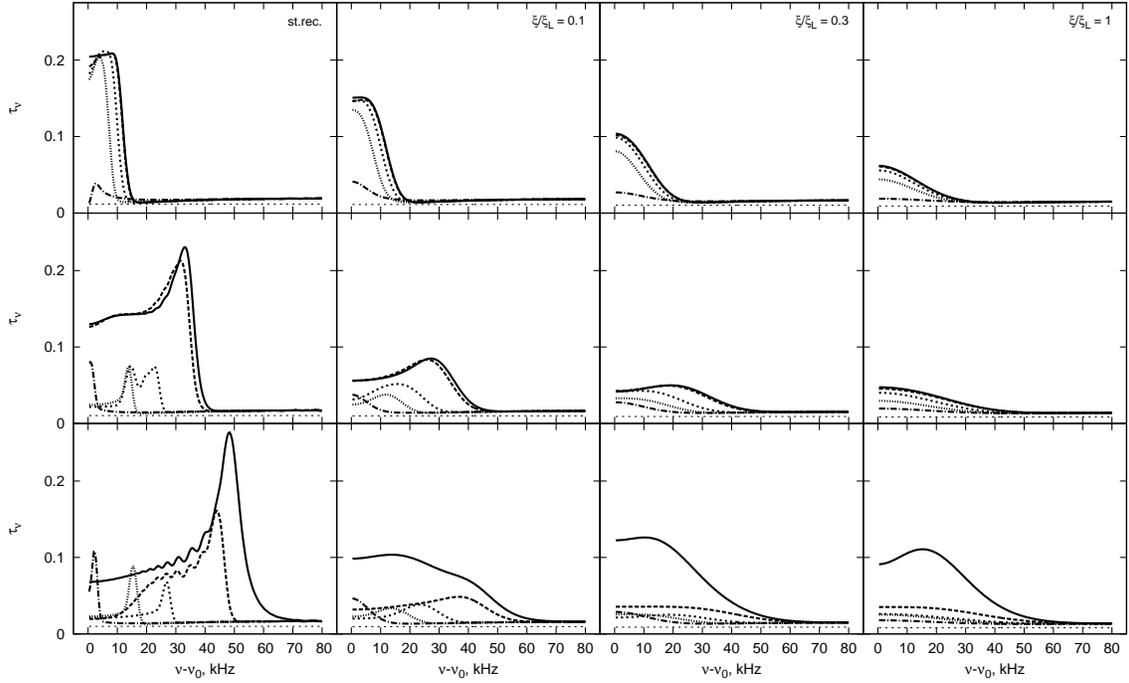}
\caption{
The same as in Figure~\ref{fig4}, but for a halo with $M=10^7\msun$: $\alpha=0.1,\  0.3,\  1,\  1.5,\ 3r/r_{vir}$ (solid, dash, dot, short dash and dot-dash lines, correspondingly), at redshifts $z = 15.5, 12$ and 10 (from top 
to bottom).
}
\label{fig6}
\end{figure*}

Figures~\ref{fig4}-\ref{fig6} show the optical depth at several impact parameters, $\alpha$, for the haloes with
$M=10^5,\  10^6,\  10^7\msun$ at redshifts $z = 15.5, 12$ and 10 (from top to bottom, correspondingly)\footnote{For
$M=10^7\msun$ the lowest redshift is $z\simeq11$, see Section~5.1 for details.} in the standard recombination model 
and in the presence of decaying dark matter with $\xi/\xi_{L} = 0.1, 0.3, 1$. The analysis of the 21 cm absorption
features of minihaloes in the standard recombination scenario can be found in \citep{ferrara11,meiksin11,vs12}. 

\subsubsection{Line profiles in on-star-forming minihaloes: $M=10^5\msun$ and $10^6\msun$.}

\begin{figure*}
\includegraphics[width=70mm]{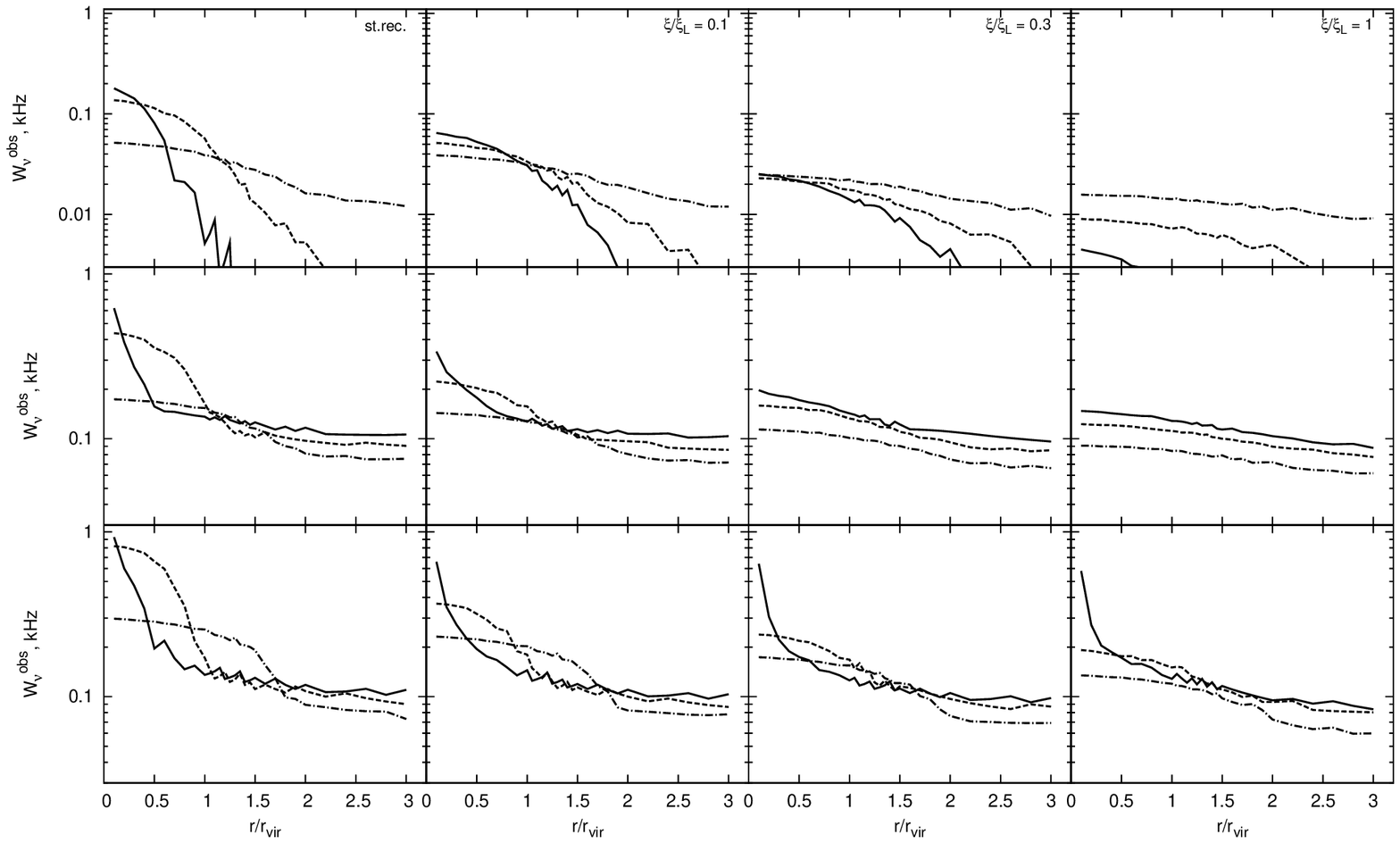}
\caption{
The observed equivalent widths for minihaloes with mass $M=10^5,\  10^6,\  10^7\msun$ (from top to bottom) at 
redshifts $z = 15.5, 12$ and $z=z_{vir}=10$ (dot-dash, dash and solid lines, correspondingly) in the standard
recombination model and in the presence of decaying dark matter with $\xi/\xi_{L} = 0.1, 0.3, 1$ (from left to 
right, correspondingly), where $\xi_{L} = 0.59\times 10^{-25}$~s$^{-1}$. Note that the y-axes are different 
between panels.
}
\label{fig7}
\end{figure*}

Minihaloes with $M=10^5~\msun$ show a small peak in optical depth at $\nu-\nu_0\simeq 40$~kHz (Figure~\ref{fig4}). 
This peak emerges from dynamics of the halo: it forms in a narrow external layer of the halo where a non-zero 
diverging velocity overlaps with a fast decreasing temperature profile -- $b^2(r)$ in the denominator of the 
exponent (\ref{optd}). The peak weakens in models with decaying DM. Note very weak or lack of dependence of the 
optical depth at $\nu-\nu_0 > 10$~kHz on the impact parameter, reflecting the fact that the velocity profiles 
are flat within haloes and do not practically depend on the ionization parameter (\ref{fig1}). In the standard 
scenario optical depth drops sharply at higher frequencies, $\nu-\nu_0 > 50$~kHz, from one side because temperature 
at the boundary falls down to very low background temperature, and from the other, because velocity at the boundary
joins the Hubble flow (such that the numerator in the exponent of (\ref{optd}) increases). In the presence of
decaying dark matter the peak at $\nu-\nu_0\simeq 40$~kHz is suppressed due to the increase of the background
temperature (leading to an increase of denominators both in the exponent and in the integrand of (\ref{optd})).
For higher masses similar peaks are found at higher frequencies, though they contribute little to absorption.

An obvious result is that the presence of decaying dark matter particles heavily suppresses optical depth:  
for $M=10^5\msun$ in the standard recombination scenario optical depth at $\alpha = 0.1$ reaches $\sim$0, 
and falls below $\sim$0.06 for $\xi > 0.1\xi_L$. For $M=10^6\msun$ optical depth shows similar dependence, though  
frequency profiles at $\alpha > 0.3$ seem in principle distinguishable at lower ionization rates $\xi \simlt 0.3\xi_L$ 
(Figure~\ref{fig5}). Thus, the decaying dark matter practically erases 21 cm absorption features from minihaloes 
with $M=10^5-10^6\msun$ and the ionization rate higher than $\xi \simgt 0.3\xi_L$.

Horn-like profiles in $M=10^6\msun$ haloes at $z=12$ (middle left panel of Figure~\ref{fig5}) are due to continuing
accretion of atomic hydrogen at these stages -- such profiles are first described by \cite{ferrara11}. They disappear 
either during further evolution when the halo settles down to a virial state 
(lower left panel), or in the models with an increase of the decaying dark matter where heating prevents contraction
(middle row of panels).

\subsubsection{Line profiles in star-forming minihaloes: $M=10^7\msun$.}

In the standard recombination horn-like profiles due to strong accretion in massive haloes around its virialization
\citep{ferrara11,vs12} are clearly seen at $\alpha \simlt 0.3$. More precisely, they origin due to 
lower spin temperature in the region with the maximum infall velocity as seen from the integrand in (\ref{optd}).
Higher infall velocity for a minihalo $M=10^7\msun$ shifts maximum of the optical depth towards higher frequencies (cf. 
the position of the optical depth maximum at $z=12$ and $z=10$ for e.g. $\alpha=0.1$, see Figure~\ref{fig3}). 

In general, in the presence of decaying dark matter the accretion rate (i.e. the infall velocity) weakens (see
Figure~\ref{fig3}), because of warming of the accreting gas. As a consequence the optical depth maximum shifts
to lower frequencies and a horn-like dependence of the optical depth almost disappears for $\xi \simgt 0.1\xi_L$.

\subsection{Equivalent width}

The suppression of the 21 cm line optical depth by decaying DM particles manifests in decrease of the observed 
line equivalent width determined as $W_\nu^{obs} =W_\nu / (1+z)$, where the intrinsic equivalent width is 
\be
 {W_\nu \over 2} = \int_{\nu_0}^{\infty}{(1-e^{-\tau_\nu})d\nu} 
                          - \int_{\nu_0}^{\infty}{(1-e^{-\tau_{IGM}})d\nu}
\label{eweq}
\ee
where $\tau_{IGM}$ is the optical depth of the background neutral IGM. Figure~\ref{fig7} demonstrates the observed
equivalent width for minihaloes with mass $M=10^5,\  10^6,\  10^7\msun$ at redshifts $z = 15.5, 12$ and 10, for the
standard recombination and in the presence of decaying dark matter with $\xi/\xi_{L} = 0.1, 0.3, 1$. As expected the
equivalent width decreases significantly with an increase of $\xi$: the equivalent width at the impact parameter 
$r/r_{vir}\simlt 0.5$ for $M=10^5\msun$ at $z=10$ (solid lines) decreases from $\sim 0.2$~KHz for the standard
recombination to a negligible value $\simlt 0.01$~kHz for $\xi/\xi_{L} = 1$. At $z = 15.5, 12$ (dash and dot-dash 
lines) the equivalent width shows not such a catastrophic drop, however it hardly can be resolved on future telescopes
and deposit into the total signal in broad-band observations \citep{ferrara11}. 

Lines with equivalent widths $W_\nu^{obs} \simgt 0.2$~kHz can be formed by haloes with $M\simgt 10^6\msun$ at 
$z=10$ (solid lines) for $\xi/\xi_{L} \simlt 0.3$ up to $\alpha = r/r_{vir}\simlt 1$. In the higher redshift range,
$z=15.5, 12$, such strong lines can be associated only with $M\simgt 10^6\msun$ in the standard recombination model.
Higher equivalent widths $\sim 0.5-0.9$~kHz can be attributed to low impact factors $\alpha\sim 0.3$ in the standard
recombination. However, more massive haloes, $M=10^7\msun$, can form strong absorptions with the width $\simgt 0.5$~kHz 
even for high ionization rate $\xi \sim \xi_L$, though only in a narrow impact factor range $\alpha \simlt 0.1$.
Thus, the decaying dark matter leads to a considerable decrease of the number of strong absorption lines. This
suppression of the 21 cm absorption is similar to what stems from the action of the X-ray background 
\citep{furla02,ferrara11}. For $\xi/\xi_{L} \simgt 0.3$ strong lines are formed only in high-mass ($M\sim 10^7\msun$)
haloes nearly along the diameter, but such events are rarely expected. 

\begin{figure}
\includegraphics[width=80mm]{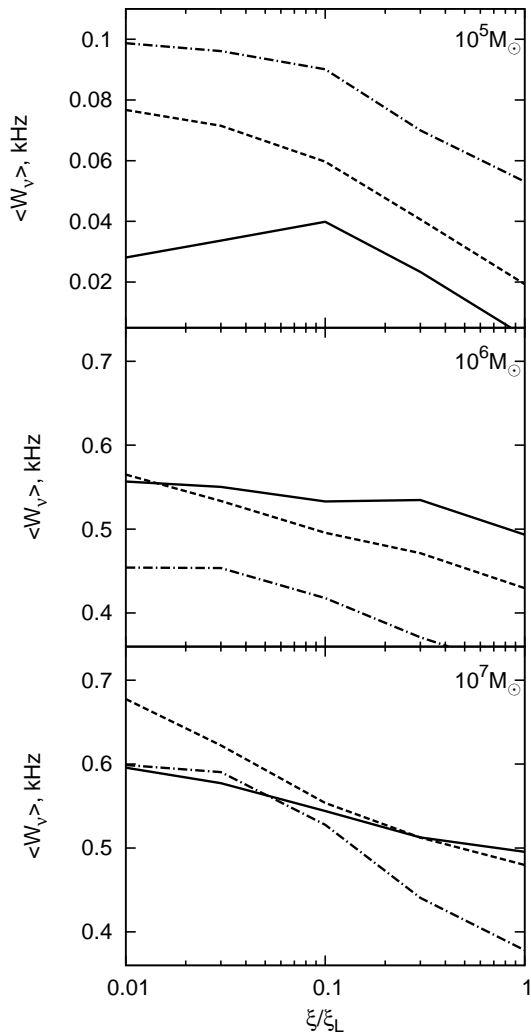}
\caption{
The dependence of the radially averaged observed equivalent width for minihaloes with mass 
$M=10^5,\  10^6\msun,\  10^7\msun$ (from top to bottom panels, correspondingly) 
at redshifts $z = 15.5, 12$ and 10 (dot-dash, dash and solid 
lines, correspondingly) on the normalized rate of the decay energy deposited in baryonic 
gas $\xi/\xi_L$, where $\xi_{L} = 0.59\times 10^{-25}$~s$^{-1}$; in the logarithmic 
$x$-axis $\xi/\xi_L = 0.01$ is set to correspond to $\xi/\xi_L = 0$, i.e. the standard recombination.
Note differences between the panels on $y$-axes scales.}
\label{fig8}
\end{figure}

The sensitivity of future telescopes is more than one order of magnitude lower than the flux limit
needed to separate spectral lines from individual haloes, i.e. to observe with spectral resolution of
1~kHz \citep{ferrara11}. Therefore, \citet{ferrara11} have proposed broad-band observations with lower
resolution. In this case one measures averaged signal from different haloes crossed by a line of sight.
Accordingly, we introduce a radially averaged observed equivalent width: $\langle W_\nu^{obs}\rangle =
\int_0^{3r_{vir}}W_\nu^{obs}(r)rdr/r_{vir}^2$. Figure~\ref{fig8} presents the dependence of this value 
for minihaloes $M=10^5,\  10^6\msun,\  10^7\msun$ on $\xi/\xi_L$. The increase of $\langle W_\nu^{obs}\rangle$ 
for $M=10^5\msun$ at $z=10$ and $\xi/\xi_L \leq 0.1$ is explained by a widening of $W_\nu^{obs}$ in the range 
of impact factor up to $r/r_{vir}\simlt 2$ (see upper panels in Figure~\ref{fig7}). For other considered masses 
and redshifts the radially averaged $W_\nu^{obs}$ decreases when $\xi/\xi_L$ grows by a factor of 1.2-2, which
apparently is sufficient to identify effects from additional ionization sources, and in particular, from decaying 
DM particles.

In previous studies theoretical spectrum of the 21 cm forest has been simulated within the assumption of 
steady-state minihaloes with fixed profiles corresponding to the virialization \citep{furla02,ferrara11}. 
When dynamics is taken into account, minihaloes lying on a given line of sight are, in general, at different 
evolutionary stages, and have different density, velocity and temperature profiles. Moreover, Press-Schechter 
formalism cannot anymore be used for description of their mass function as soon as they are lying far from the
virialization. It makes problematic to describe correctly the number density of minihaloes at each evolutionary 
stage. Possible formation of stars in massive evolved minihaloes brings additional complication. Overall, the 
analysis of theoretical spectrum within a statistical simulation becomes exceedingly cumbersome. More relevant 
picture can be obtained from high-resolution cosmological gas dynamic simulation, although current resolution 
with $\Delta M\sim 10^6~M_\odot$ seems not to be sufficient.

However, with the upcoming low frequency interferometers (LOFAR and SKA) such modeling seems to be excessive and
unnecessary. Indeed, the absorption line profiles and the spectral features from separate minihaloes located at 
$z=10$ are of 1 to 5 kHz in width -- close to the resolution limit 1 kHz. As mentioned \cite{ferrara11} the 
sensitivity corresponding to such a resolution requires enormously bright background sources, GRB afterglows 
and/or QSOs, and broad-band observations can be a better alternative. In broadband observations the suppression 
of optical depth in 21 cm caused by decaying particles manifests as a factor of 2--4 decrease of absorption in 
the frequency range $\nu<140$ MHz where contribution from low mass minihaloes at $z\geq 10$ dominates. Indeed, 
the number of haloes in the low-mass range $10^5-10^7\msun$ at $z=10$ scales as $n(M,z)\sim M^{-1}$ (the 
Press-Schechter mass function). The probability of minihaloes to intersect a line of sight is proportional 
to $n(M,z)\times(\alpha r_{vir}(M))^2$, resulting in a decrease of the number of strong absorption lines 
with $W_\nu^{obs} \simgt 0.3$~kHz by factor of at least 2.5 for $\xi/\xi_{L} = 0.3$ and more than 4.5 for 
$\xi/\xi_{L} = 1$. Such a decrease inevitably cancels the average 21 cm forest signal from low mass haloes 
expected in future broadband observations.

\section{Conclusions}

In this paper we have considered the influence of decaying dark matter particles on the HI 21 cm absorption 
features from low mass minihaloes with mass $M=10^5,\  10^6\msun,\  10^7\msun$ virialized at $z_{vir} = 10$. 
We used a 1D self-consistent hydrodynamic approach to study their evolution, and followed through the absorption
characteristics from the turnaround to the virialization of minihaloes. 
We have found that

\begin{itemize}
\item due to an additional heating brought by decaying particles thermal and dynamical evolution of minihaloes 
in the presence of decaying dark matter shows cosiderable difference from that in the model without particle 
decays (i.e. with the standard recombination scenario); 

\item this additional heating strongly suppresses optical depth in the 21 cm line: it practically ``erases'' 
the 21 cm absorptions from minihaloes with $M=10^5-10^6\msun$ even at a relatively modest ($\xi \simgt 0.3\xi_L$)
ionization rate from decaying DM particles; the horn-like dependence of the optical depth found for minihaloes  
with $M \sim 10^6 - 10^7\msun$ in the standard recombination scenario \cite{ferrara11} almost disappears even 
at a lower ionization rate $\xi \simgt 0.1\xi_L$;

\item the equivalent width of the 21 cm absorption line decreases significantly while $\xi$ increases, and 
the number of strong absorption lines $W_\nu^{obs} \simgt 0.3$~kHz at $z=10$ drops by more than 2.5 to 4.5 times
depending on $\xi$; such a decrease inevitably erases the averaged 21 cm forest signal from low mass haloes 
($M \sim 10^6 - 10^7\msun$, i.e. the frequency range $\nu<140$ MHz) in future broad-band observations.
\end{itemize}

\section{Acknowledgements}

This work is supported by the Federal Agency of Education (project codes RNP 2.1.1/11879, P-685).
EV acknowledges support from the "Dynasty" foundation.



\end{document}